\documentclass[aps,prl,reprint,groupedaddress,amsmath,amssymb]{revtex4-1}

\usepackage[utf8]{inputenc}
\usepackage{bpchem}
\usepackage{physics}
\usepackage{bbm}
\usepackage{graphicx}
\usepackage[usenames,dvipsnames,table]{xcolor}
\usepackage{soul}
\usepackage[caption=false]{subfig}
\usepackage{hyperref}
\usepackage{cleveref}

\DeclareMathOperator{\polylog}{Li}

\crefformat{figure}{#2Fig.~#1#3}

\begin{document}


\title{Tilting the balance towards $d$-wave in iron-based superconductors}
\author{Mario Fink}
\author{Ronny Thomale}
\email{rthomale@physik.uni-wuerzburg.de}
\affiliation{Institut für Theoretische Physik, Universität Würzburg,
  Am Hubland, D-97074 Würzburg, Germany}
\date{\today}

\begin{abstract}
The intricate interplay of interactions and Fermiology can give
rise to a close
competition between nodeless (e.g. $s$-wave) and nodal (e.g. $d$-wave) order in
electronically driven unconventional superconductors. We analyze how
such a scenario is affected by a Zeeman magnetic field $H_{\text{Z}}$ and temperature $T$. In the neighborhood of a zero temperature first order
critical point separating a nodal from a nodeless phase, the phase
boundary at low $H_{\text{Z}}$ and/or low $T$ has a universal line shape cubic in
$H_{\text{Z}}$ or $T$, such that the
nodal state is stabilized at the expense of the nodeless. We
calculate this line shape for a model of competing $s_\pm$-wave and
$d$-wave pairing in iron-based superconductors.
\end{abstract}


\maketitle


{\it Introduction.} Superconductivity is a quantum many-body state of matter which is a phase-coherent
superposition of paired electrons \cite{schrieffer1983theory}. In order to
estimate whether such a state is energetically preferable, we have to consider
both the aspects of energy gain due to the formation of the pair condensate
as well as the conditions of pairing, i.e., the challenge to overcome the Coulomb
repulsion between the electrons. In phonon-mediated superconductors, the
interplay of electrons and lattice distortions imply an effective attraction
between electrons at appropriate retardation scales \cite{PhysRev.108.1175}; the
generically preferred superconducting phase then is an $s$-wave state, which optimizes the
condensation energy and, at the same time, is least susceptible to disorder. For 
electronically driven superconductors, the initial challenge is to avoid Coulomb
repulsion, which, despite the screening of the long range part of the
interactions due to the electron fluid, is still present at small
distances.
In the cuprates, this is accomplished
by $d$-wave superconductivity, i.e., through establishing a condensate with
finite relative angular momentum of the Cooper pairs. Depending on an itinerant or localized moment
point of view, such a condensate is either preferred due to an enhanced $(\pi,\pi)$ channel
of particle hole fluctuations \cite{PhysRevLett.15.524} or due to doping an
antiferromagnetic Mott insulator \cite{ANDERSON1196,RevModPhys.78.17}. Condensation
energy dictates which of the possible $d$-wave solutions
is preferred: among the in-plane polarized states, the $d_{x^2-y^2}$-wave solution is chosen over $d_{xy}$-wave, as
the Fermi level density of states of the cuprate band structure is smallest along the nodal
lines of $d_{x^2-y^2}$, which guarantees a minimal loss of
condensation energy by the presence of the nodes. Due to the momentum dependence of pairing, $d$-wave is more
susceptible to disorder than $s$-wave~\cite{andi-dis}.

Iron-based superconductors \cite{doi:10.1021/ja800073m}  are located such in
parameter space that maximizing condensation energy and minimizing Coulomb
repulsion are of similar importance. Despite of electron-electron interactions
as the apparent dominant role in pairing \cite{PhysRevLett.101.026403}, the
interaction scales are weaker than in the cuprates, and multiple Fermi
surfaces as well as multiple orbitals that contribute to the pairing significantly
complicate the picture \cite{0034-4885-74-12-124508,doi:10.1146/annurev-conmatphys-020911-125055}. Different theoretical
approaches from an itinerant
\cite{PhysRevLett.101.087004,PhysRevB.80.180505,2009NJPh...11b5016G} and
localized moment \cite{PhysRevLett.101.076401,PhysRevLett.101.206404} picture
have found $s$-wave and $d$-wave to be in close competition to each other for
iron-based superconductors.
The majority of experimental evidence suggests $s$-wave superconductivity for
most pnictide compounds \cite{RevModPhys.83.1589}, however, this observation might
change as the crystal quality will improve in upcoming waves of
refined material synthesis.
Some indication along these lines has been
recently obtained in \BPChem{KFe\_2As\_2} as the most strongly
hole-doped limit of \BPChem{K\_{x}Ba\_{1-x}Fe\_2As\_2}, where the
crystal quality is significantly enhanced through the absence of any Ba
content. There, as predicted theoretically~\cite{PhysRevLett.107.117001}, thermal conductivity measurements
\cite{2012PhRvL.109h7001R,PhysRevB.89.064510} have found strong indication for
$d$-wave, while the overall situation is still far from
settled (see e.g. Ref.~\onlinecite{PSSB:PSSB201600350} and references therein).
Recently, Raman scattering has found $d$-wave pairing to be
of nearly competitive propensity to $s$-wave in
\BPChem{K\_{x}Ba\_{1-x}Fe\_2As\_2} crystals of low to intermediate hole doping
level $0.22<x<0.7$~\cite{tommy}. The specific details of this
subleading $d$-wave state fit the superconducting form factor
predicted in~\cite{PhysRevB.80.180505}.

In this Letter, we elaborate on how the competition between an $s$-wave and
a $d$-wave state is affected perturbatively by a Zeeman field $H_{\text{Z}}$ or
temperature $T$. As a phenomenological starting point, we place ourselves
at a first order zero temperature critical point where we assume a gapped $s$-wave and a nodal
$d$-wave state to be of equal energy density. As partly elaborated on
above, such a situation might occur in a multiple Fermi surface
scenario with a balanced energetic significance of pairing formation and
condensation energy.
When a weak Zeeman field is turned on, the
$s$-wave state will not respond to it since all quasiparticles are gapped. The
$d$-wave state, however, possesses gapless quasiparticles at the nodes,
which are polarized due to the Zeeman field.
This gain in magnetic polarization energy makes the $d$-wave state
preferable, and thus serves as a parameter to tune a phase transition
from $s$-wave to $d$-wave. This similarly applies for finite
temperature, where the $d$-wave state, in contrast to the $s$-wave
state, gains free energy through the generation of entropy in the
nodal regime. It warrants similarity to the Pomeranchuk effect in He$^3$
where the crystal phase is stabilized at finite $T$~\cite{pome,rich}. 


{\it Model.} At zero temperature, a singlet superconductor subject to a Zeeman field is
described by the Hamiltonian
\begin{eqnarray}
		H
	=
		\sum_{{\bf k},\sigma,\sigma'}
		\left(
			\varepsilon({\bf k})
		-
			\mu_{\text{B}} H_{\text{Z}} \sigma^{z}_{\sigma,\sigma'}
		\right)
		c^{\dagger}_{{\bf k},\sigma}
		c_{{\bf k},\sigma'}^{\phantom{\dagger}} \nonumber \\
	+
		\frac{1}{2}
		\sum_{{\bf k},\sigma,\sigma'}
		\left(
			\Delta_{\sigma,\sigma'}({\bf k})
			c^{\dagger}_{{\bf k},\sigma}
			c^{\dagger}_{-{\bf k},\sigma'}
		+
			\mathrm{h.c.}
		\right)
	\quad \text{,}
	\label{eqn:zeeman_field_hamiltonian}
\end{eqnarray}
where $\varepsilon({\bf k})$ is the kinetic term of the electrons,
$\Delta_{\sigma,\sigma'}({\bf k})=i\Delta({\bf k})\sigma^{y}_{\sigma,\sigma'}$,
and $\sigma^{y}$, $\sigma^{z}$ denote the Pauli matrices. The orbital
magnetic field contribution be
weak in comparison to the Zeeman term, e.g. as accomplished by an
in-plane field~\cite{PhysRevB.57.8566} in a quasi two-dimensional crystal~\footnote{Orbital
  magnetic field effects of a nodal superconductor generically dominate over the Zeeman term at weak field strength for a
  field orientation perpendicular to the 2D plane. The Volovik effect~\cite{volo} then
  predicts a $\propto \sqrt{H}$ scaling of the induced density of states, as
  opposed to the $\propto H$ scaling for the Zeeman term.}. For $H_{\text{Z}}=0$ and $T=0$,
our phenomenological ansatz
assumes energy densities $e_s=e_d$ of an $s$-wave and a $d$-wave
state.

\begin{figure}[t]
	\begin{center}
		\subfloat[\label{fig:zeeman_field_spm}]{
      \includegraphics{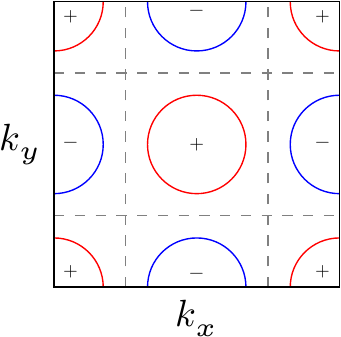}
		}
    \hspace{0.0cm}
		\subfloat[\label{fig:zeeman_field_dpm}]{
      \includegraphics{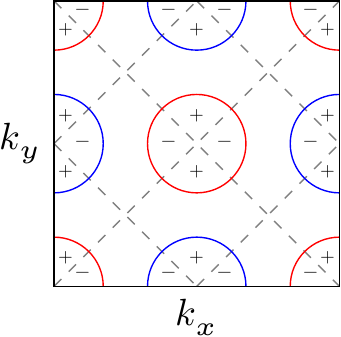}
		}
\caption{Schematic plot of the (a) $s^\pm$-wave and (b) $d^\pm$-wave superconducting
  gap function for typical electron (blue) and hole pockets (red) in
  the iron pnictides. "+'' and "-'' label the sign of the gap
  functions in the different domains of the Brillouin zone. The zeroes of the gap function (grey dashed lines) do not
  cross the Fermi surfaces in (a), but yield four nodes for each Fermi
  pocket in (b). The $s\pm$ gap function in (a) corresponds to a second
  nearest neighbor function of the $A_{1g}$ representation $\Delta^{s\pm}({\bf k})\propto\cos(k_x)\cos(k_y)$.  The $d\pm$ gap
  function in (b) corresponds to a third nearest neighbor function of
  the $B_{1g}$ representation,
  $\Delta^{d\pm}({\bf k})\propto\cos(2k_x)-\cos(2k_y)$~\cite{PhysRevB.80.180505}.
  }
	\label{fig:zeeman_field_gap_symmetries}
	\end{center}
\end{figure}

As a specific competing candidate state in materials such as the pnictides,
we consider extended s-wave (s$_{\pm}$)
$\Delta^{s_{\pm}}({\bf k})= \Delta_{s}\left( \cos(k_x) \cos(k_y)
\right)$~\cite{PhysRevLett.101.057003,PhysRevLett.102.047005} and
extended $d$-wave $\Delta^{d\pm}({\bf k})= \Delta_{d}\left ( \cos(2k_x)
  -\cos(2k_y) \right)$~\cite{PhysRevB.80.180505}
(\cref{fig:zeeman_field_gap_symmetries}). This choice is motived by recent
Raman scattering experiments on K-doped Ba-122 where both the leading
and subleading superconducting order can be detected, rendering the
$s_\pm$-wave and $d_\pm$-wave states close competitors~\cite{tommy}.
In our argument to follow, it is irrelevant which type of $d$-wave or $s$-wave is realized. The only
assumption is that the anisotropy in the $s$-wave state be small enough to
provide a minimal quasiparticle gap $\Delta_s > \mu_{\text{B}} H_{\text{Z}}$ (the
reduction of $s^\pm$-wave gap anisotropy, which by itself can be vital to
explaining the experimental  data in iron pnictides~\cite{PhysRevLett.106.187003}, is often caused by disorder~\cite{PhysRevB.79.094512}), and that the
$d$-wave state is nodal. The latter is a property protected by symmetry, as long as
the nodal lines intersect with at least one Fermi pocket. \\





\begin{figure}[t]
	\begin{center}
    \subfloat[\label{fig:dwave_node_define_momenta}]{
      \includegraphics[]{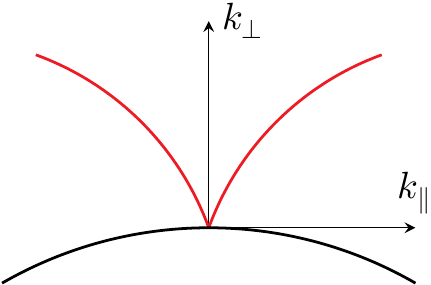}
		}
		\subfloat[\label{fig:quasiparticle_spectrum_dwave}]{
      \includegraphics[]{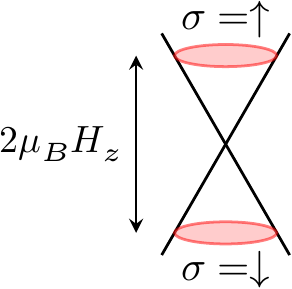}
		}
	\caption{(a) In the vicinity of a node, the $d$-wave spectrum features a linear
					 momentum dependence characterized by $k_{\perp}$ and $k_{\parallel}$.
					 (b) In the presence of a finite magnetic field, the point-like Fermi surface
           of the Bogliobov cone splits into two elliptic Fermi surfaces for opposite spins.}
	\label{fig:node_momenta_zeeman_shift}
	\end{center}
\end{figure}

{\it Bogoliubov quasiparticles.} To obtain the SC quasiparticle spectrum, we represent
(\ref{eqn:zeeman_field_hamiltonian}) by its Bogoliubov-de Gennes (BdG) form,
i.e. we the employ Nambu spinor
$\vec{C}^{\dagger}_{{\bf k}}=\left(c^{\dagger}_{{\bf k},\uparrow},c^{\dagger}_{{\bf k},\downarrow},
                                   c_{-{\bf k},\uparrow}^{\phantom{\dagger}},c_{-{\bf k}^{\phantom{\dagger}},\downarrow}\right)$
notation and obtain


\begin{equation}
		\mathcal{H}
	=
		\frac{1}{2}
		\sum_{{\bf k}}
			\vec{C}^{\dagger}_{{\bf k}}
			\left(\begin{matrix}
				\varepsilon({\bf k})-\mu_{\text{B}} H_{\text{Z}} \sigma^{z} & i\Delta({\bf k})\sigma^{y} \\
				-i\Delta^{*}({\bf k})\sigma^{y} & -\varepsilon({\bf k})+\mu_{\text{B}} H_{\text{Z}} \sigma^{z} \\
			\end{matrix}\right)
			\vec{C}_{{\bf k}}\text{.}
	\label{eqn:bogoliubov_quasiparticles_hamiltonian}
\end{equation}
The eigenvalue spectrum of
(\ref{eqn:bogoliubov_quasiparticles_hamiltonian}) is given by
\begin{eqnarray}
		E^{\pm,\pm}({\bf k})
	=
		\pm \mu_{\text{B}} H_{\text{Z}}
		\pm \sqrt{ \varepsilon({\bf k})^2 + \abs{\Delta({\bf k})}^2}\text{,}
	\label{eqn:bogoliubov_quasiparticle_spectrum}
\end{eqnarray}
where the two binary indices label the up/down spin (later denoted $\sigma$) and Bogoliubov
particle/hole character, respectively.  The eigenstates are given by
\begin{eqnarray}
		\psi^{\dagger,(-,\pm)}_{{\bf k}}
	&=&
		\Delta({\bf k})
		c^{\dagger}_{{\bf k},\uparrow}
	-
		\left(\varepsilon({\bf k})\mp\sqrt{\varepsilon^2({\bf k})+\abs{\Delta({\bf k})}^2} \right)
		c_{-{\bf k},\downarrow}
	\nonumber \\
		\psi^{\dagger,(+,\pm)}_{{\bf k}}
	&=&
		\Delta({\bf k})
		c^{\dagger}_{{\bf k},\downarrow}
	+
		\left(\varepsilon({\bf k})\mp\sqrt{\varepsilon^2({\bf k})+\abs{\Delta({\bf k})}^2} \right)
		c_{-{\bf k},\uparrow}
 . \nonumber \\
	\label{eqn:bogoliubov_quasiparticle_eigenstates}
\end{eqnarray}
\begin{figure}[t]
	\begin{center}
		\subfloat[\label{fig:quasiparticle_spectrum_swave}]{
      \includegraphics{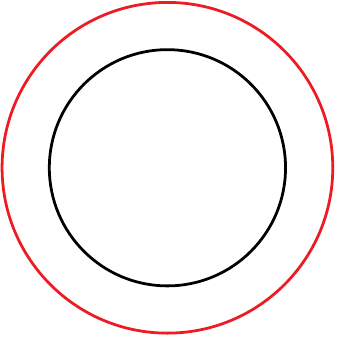}
		}
    \hspace{0.5cm}
		\subfloat[\label{fig:quasiparticle_spectrum_dirac_cone}]{
      \includegraphics{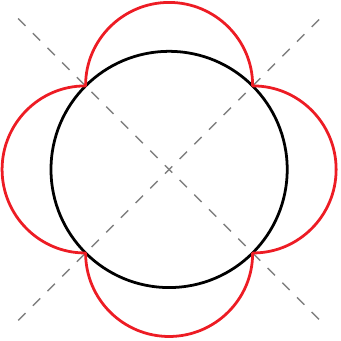}
		}
    \caption{(a) Schematic representation of the quasiparticle spectral gap
             (solid red line)
             (\ref{eqn:bogoliubov_quasiparticle_spectrum}) for an
             ideal $s$-wave scenario $\Delta({\bf k})=\Delta_{s}$ displaying a
             finite gap along the entire Fermi surface (solid black line).
             (b) In contrast, the $d$-wave gap exhibits nodes (dashed lines)
             where the gap vanishes.
           }
    \label{fig:quasiparticle_spectra}
	\end{center}
\end{figure}
{\it Quasiparticle magnetization.} Assume $k_{\text{B}} T \approx 0$ to be the smallest of all energy scales, and
consider the spectrum~(\ref{eqn:bogoliubov_quasiparticle_spectrum}) in the case
of $\Delta({\bf k})=\Delta^{s_\pm,d_\pm} ({\bf k})$ for one representative Fermi pocket
(\cref{fig:quasiparticle_spectra}): for the case of $s_\pm$-wave in
\cref{fig:quasiparticle_spectrum_swave}, $\Delta({\bf k}) \approx \Delta_s$.
For $\mu_{\text{B}} H_{\text{Z}} < \Delta_s$, the Zeeman field has a negligible effect on the
spectrum because the binding energy of the singlet pair is larger than the
applied field. For $d_\pm$-wave, however, while gapped by
$\Delta({\bf k}) \approx \Delta_d$ in the anti-nodal regime, there is a nodal
regime $\Delta({\bf k}) \approx 0$ illustrated in
\cref{fig:quasiparticle_spectrum_dwave} where the quasiparticles will respond to the
Zeeman field~\cite{PhysRevB.57.8566}. 
We expand the quasiparticle dispersion (\ref{eqn:bogoliubov_quasiparticle_spectrum})
around the nodal regime in momentum space at ${\bf k}^{*}_{\text{F}}$. Define
\begin{eqnarray}
		v_{\text{F}}
	\equiv
		\frac{\partial E({\bf k})}{\partial k_{\perp}}
		\bigg|_{{\bf k}={\bf k}^{*}_{\text{F}}}
	\quad \text{,} \quad
		v_{\Delta}
	\equiv
		\frac{\partial E({\bf k})}{\partial k_{\parallel}}
		\bigg|_{{\bf k}={\bf k}^{*}_{\text{F}}}
	\quad \text{,}
	\label{eqn:quasiparticle_magnetization_fermi_velocities}
\end{eqnarray}
where $k_{\perp}$ and $k_{\parallel}$ are the momenta perpendicular and
tangential to the Fermi surface of $\varepsilon({\bf k})$
(\cref{fig:dwave_node_define_momenta}).
The dispersion takes the form of an elliptic Dirac cone
\begin{eqnarray}
		E({\bf k}) \big|_{{\bf k}\approx{\bf k}^{*}_{\text{F}}}
	=
		v_{\text{F}} k_{\perp} + v_{\Delta} k_{\parallel}
	\quad \text{.}
	\label{eqn:quasiparticle_magnetization_dirac_cone}
\end{eqnarray}
Due to $H_{\text{Z}}$, the Fermi levels of spin $\uparrow$ and $\downarrow$
quasiparticles shift against each other, and a Fermi pocket emerges, with its
surface given by the elliptic equation
\begin{eqnarray}
		\left(k_{\perp}-k^{*}_{F,\perp}\right)^{2}
		v_{\text{F}}^{2}
	+
		\left(k_{\parallel}-k^{*}_{F,\parallel}\right)^2
		v_{\Delta}^{2}
	=
		\mu_{\text{B}}^{2} H_{\text{Z}}^{2}
	\quad \text{.}
	\label{eqn:quasiparticle_magnetization_elliptic}
\end{eqnarray}
Computing the magnetization from there, without loss of generality, we constrain ourselves
to the spin $\uparrow$ species. The number of particles in a given volume
$\mathcal{V}$ is given by the area enclosed by the Fermi surface
(\ref{eqn:quasiparticle_magnetization_elliptic})
\begin{eqnarray}
		N_{\uparrow}
	&=
		\matrixelement{FS}{n_{{\bf k},\uparrow}}{FS}
	\nonumber \\
	&=
		\frac{\mathcal{V}}{(2\pi)^{2}} \pi \frac{\mu_{\text{B}}^{2}H_{\text{Z}}^{2}}{v_{\text{F}}v_{\Delta}}
	\quad \text{.}
	\label{eqn:quasiparticle_magnetization_number_volume}
\end{eqnarray}
From the density $n_{\uparrow}(E)=\frac{1}{4\pi}\frac{E^2}{v_{\text{F}} v_{\Delta}}$, we
obtain the energy density of states
$\rho(E)=\frac{\partial n_{\uparrow}}{\partial E}=\frac{1}{2\pi}\frac{E}{v_{\text{F}} v_{\Delta}}$.
Considering both spin $\uparrow$ and $\downarrow$ contributions, we
eventually obtain the
quasiparticle magnetization density
\begin{eqnarray}
		m
	=
		\mu_{\text{B}} \left( n_{\uparrow} - n_{\downarrow} \right)
	&=
		2 \mu_{\text{B}}
		\int_{0}^{\mu_{\text{B}}H_{\text{Z}}} \ \rho(E) \ \mathrm{d}E
	\nonumber \\
	&=
		\frac{\mu_{\text{B}}}{2\pi} \frac{\mu_{\text{B}}^{2}H_{\text{Z}}^{2}}{v_{\text{F}}v_{\Delta}}
	\quad \text{.}
	\label{eqn:quasiparticle_magnetization_density}
\end{eqnarray}
Assuming we had started from an equal energy density for $s$-wave and $d$-wave, the
Zeeman field hence yields a \emph{preference} for $d$-wave according to
\begin{eqnarray}
		e_d - e_s
	=
		-m H_{\text{Z}}
	=
		-N_{n}
		\frac{\left(\mu_{\text{B}} H_{\text{Z}}\right)^{3}}{2\pi v_{\text{F}} v_{\Delta}}
	\quad \text{,}
	\label{eqn:quasiparticle_magnetization_preference}
\end{eqnarray}
where $N_n$ denotes the total number of nodes induced by the $d$-wave state on the
multi-pocket Fermi surface. \\

\begin{figure}[t]
      \includegraphics{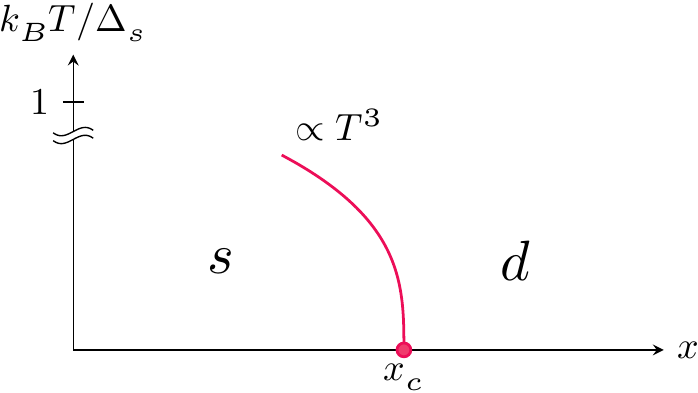}
\caption{Sketch of the phase boundary between a competing gapped $s$-wave and
nodal $d$-wave state as a function of tuning parameter $x$ and
temperature $T$. Emanating from a first order quantum critical point
at $x_c$,
there is a universal cubic trajectory for temperatures below the gap
scale $\Delta_s$.}
	\label{phasediagram}
\end{figure}

{\it Quasiparticle entropy.}
Assume $H_{\text{Z}}\approx 0$ and finite temperature $k_{\text{B}}T < \Delta_{s}$.
The free energy density is given by
$		f
	=
		u - Ts
	=
		-k_{\text{B}} T/N \ln \mathcal{Z}$,
where $u$ denotes the inner energy density, $T$ the temperature, $s$ the
entropy per particle, and $\mathcal{Z}$
the (grand) canonical partition function. The latter be denoted by
\begin{eqnarray}
		\mathcal{Z}
	=
		\sum_{\{\ket{\Psi}\}}
			e^{-\beta E_{\ket{\Psi}}}
	=
		\prod_{{\bf k}\sigma}
			\left(
				1 + e^{-\beta E^{\sigma,+}({\bf k})}
			\right)
	\quad \text{,}
	\label{eqn:entropic_transition_partition_function}
\end{eqnarray}
with $\frac{1}{\beta}=k_{\text{B}}T$, spin index $\sigma=\pm$, many-particle state $\ket{\Psi}$, and the
corresponding energy $E_{\ket{\Psi}}$ constructed from the single-particle states
(\ref{eqn:bogoliubov_quasiparticle_eigenstates}) and the quasiparticle energy
$E^{\sigma,+}({\bf k})$ given by
(\ref{eqn:bogoliubov_quasiparticle_spectrum}), as we only consider the
particle-type Bogoliubov branch. We set $e_d=e_s$ such that for the
relative change in free energy density $f_d-f_s$, the inner
energy does not enter.
The nodal regime of the $d$-wave state then is the only relevant
contribution to the change of $f_d-f_s$ at low temperatures. Since $f$ depends only on the momentum ${\bf k}$ through the energy
$E({\bf k})$, we express it by means of the density of states
$\rho_{\sigma,+}(E({\bf k}))$
to find
\begin{eqnarray}
		f
	=
		-k_{\text{B}}T
		\sum_{\sigma}
		\int_{-\infty}^{+\infty}
		\hspace{-6pt}	\rho_{\sigma,+}(E)
			\ln\left(1+e^{-\beta E^{\sigma,+}}\right)
		\mathrm{d}E.
  \label{eqn:free_energy_density_of_states}
\end{eqnarray}
Due to the discontinuity of $\rho_{\sigma,+}(E)$ at the band edge,
we have to include the boundary terms upon partial integration. The
boundary terms, however, may be neglected for $\beta\gg1$, which,
together with $f_s \approx 0$ in this limit, leaves us with
\begin{equation}
		f_d-f_s
	\overset{\beta \gg 1}{\approx}
		-
		\sum_{\sigma}
		\int_{0}^{E_{\text{max}}}
			\frac{\Upsilon_{\sigma,+}(E)}{e^{\beta E^{\sigma,+}}+1}
		\mathrm{d}E
	\quad \text{,}
	\label{eqn:entropic_transition_free_energy_integration_by_parts}
\end{equation}
where $\rho_{\sigma,+}(E)=\frac{\mathrm{d}\Upsilon_{\sigma,+}(E)}{\mathrm{d}E}$
and $E_{\text{max}}=\limsup_{{\bf k}\in \text{BZ}}E^{\sigma,+}({\bf k})$. We
are interested in the low-temperature behavior of $f_d$, which is why
we are allowed to model the density of states by a linear function
along with (\ref{eqn:quasiparticle_magnetization_dirac_cone}), i.e.
$\rho_{\sigma,+}(E)=\alpha E_{\sigma,+}$ such that
$\Upsilon_{\sigma,+}(E)=\frac{\alpha}{2} E_{\sigma,+}^{2}$.
%
This reduces our task to solving the Fermi-Dirac integral
\begin{equation}
		\frac{\alpha}{2}
		\int_{0}^{E_{\text{max}}}
			\frac{E^{2}}{e^{\beta E}+1}
		\mathrm{d}E=
		\frac{\alpha}{2}
		\frac{1}{\beta^{3}}
		\int_{0}^{\beta E_{\text{max}}}
			\frac{x^2}{e^{x}+1}
		\mathrm{d}x
	\quad \text{.}
	\label{eqn:entropic_transition_fermi_dirac_integral}
\end{equation}
It already shows that
the change in free energy density has the low-temperature behavior
$f_d-f_s \propto -T^{3}$. The integral is solved by
the integrand's primitive~\cite{1995JMP....36.1217L,2009arXiv0909.3653M}
\begin{eqnarray}
		\Xi(x)
	=
		\frac{x^3}{3}
	-
		x^2 \ln\left(1+e^{x}\right)
	-
		2x \polylog_{2}(-e^{x})
	+
		2 \polylog_{3}(-e^{x}) \text{,} \nonumber \\
	\label{eqn:appendix_entropic_transition_fermi_dirac_integral_primitive}
\end{eqnarray}
where $\polylog_i(x)$ denotes the polylogarithm \cite{abramowitz1972handbook}.
%
%
The result may be easily checked  using
$\frac{\mathrm{d}}{\mathrm{d}x}\polylog_{s}(x)=\frac{1}{x}\polylog_{s-1}(x)$
and $\polylog_{1}(x)=-\ln(1-x)$. The integral in~(\ref{eqn:entropic_transition_fermi_dirac_integral}) gives
\begin{eqnarray}
		\Xi(\beta E_{\text{max}})
	-
		\Xi(0)
	\approx
		- 2 \polylog_{3}(-1)
	=
		2\eta(3)
	\quad \text{,}
  \label{eqn:appendix_fermi_dirac_primitive}
\end{eqnarray}
where we neglect $\Xi(\beta E_{\text{max}})$ for small temperatures since
$\lim_{x\rightarrow\infty}\Xi(x)=0$.
We express the
polylogarithm in terms of the Dirichlet $\eta$-function, which in turn
is written in terms of the Riemann $\zeta$-function $\eta(s)=(1-2^{1-s})\zeta(s)$.
Having thus fixed the exact prefactor, we find
%
%
%
%
%
\begin{equation}
		f_d-f_s	\overset{\beta\gg 1}{\approx}
    -  \frac{3}{2} \alpha \zeta(3)  \left(k_{\text{B}}T\right)^{3}
	\quad \text{,}
	\label{eqn:entropic_transition_free_energy_low_temperature_sol}
\end{equation}
with $\zeta(3)\approx1.20206$ (Apéry's constant), and the universal
cubic phase boundary shape depicted in Fig.~\ref{phasediagram}.
 This finding is
calculated for one single node, and would have to be multiplied by
$N_n$ to account for the total number of nodes.
In~\eqref{eqn:entropic_transition_free_energy_low_temperature_sol},
$\alpha$ is the non-universal coefficient which is sensitive to the
details of the microscopic system. To calculate an explicit example, we take the one-band
Hubbard model $\epsilon({\bf k})=-2t(\cos(k_x)+\cos(k_y))-\mu$ with a nearly
circular Fermi pocket for $t=1.0$, $\mu=-1.9$. Its density of states at the
nodes (which may be derived by means of $v_{\text{F}}$ and $v_{\Delta}$ in
Eq.~\ref{eqn:quasiparticle_magnetization_fermi_velocities}) gives
$\alpha\approx 1.58$. 
{\it Conclusion.} At the latest since the discovery
of the iron pnictides,
materials with competitive unconventional nodal and nodeless
superconducting pairing
tendencies have established an intricate problem to be further studied experimentally and
theoretically. We have shown that the perturbative Zeeman-magnetic and entropic response of
nodal quasiparticles pave the way to tilt the thermodynamic balance
in favor of the nodal, in our case $d$-wave, superconducting
state by a term which is cubic in the perturbation parameter,
i.e. either $H_{\text{Z}}$ or $T$. In the neighborhood of a zero temperature
first order critical point between the nodal and the nodeless state,
this gives rise to a universal cubic phase transition line shape in
favor of the nodal state.  Whether this contribution can yield an observable magnetically or entropically induced superconducting
phase transition crucially relies on a careful material
design as well as the detailed tunability of system parameters towards
such a critical point. From a
broader perspective, iron-based superconductors are not the only class
of materials displaying comparable superconducting pairing tendencies
towards a nodal and a nodeless state. For instance, the nodeless
chiral $p$-wave state predicted for strontium ruthenate~\cite{0953-8984-7-47-002} is challenged
by a nodal $d$-wave state, where unfortunately the possibly strongly
anisotropic gap profile makes it hard to discriminate between both
superconducting states~\cite{0295-5075-104-1-17013,elena}. It is
likely, however, that the $d$-wave state eventually has to win beyond a critical
amount of strain~\cite{Hicks283}, rendering the latter an ideal experimental tuning parameter towards
the critical point regime between a nodal and nodeless superconducting
state we have envisioned in this work.

\begin{acknowledgments}
We thank S. A. Kivelson, A.~V. Chubukov, P. Hirschfeld, A.~Mackenzie, I.~I. Mazin, S.~Raghu,
and D.~Scalapino for stimulating discussions. This work was supported
by DFG-SPP 1458, DFG-SFB 1170 (project B04), and ERC-StG-336012-TOPOLECTRICS.
\end{acknowledgments}


\bibliographystyle{prsty}
\bibliography{references}

\begin{thebibliography}{10}

\bibitem{schrieffer1983theory}
J.~R. Schrieffer, {\em Theory of {S}uperconductivity} (Benjamin/Addison Wesley,
  New York, 1964).

\bibitem{PhysRev.108.1175}
J. Bardeen, L.~N. Cooper, and J.~R. Schrieffer, Phys. Rev. {\bf 108},  1175
  (1957).

\bibitem{PhysRevLett.15.524}
W. Kohn and J.~M. Luttinger, Phys. Rev. Lett. {\bf 15},  524  (1965).

\bibitem{ANDERSON1196}
P.~W. Anderson, Science {\bf 235},  1196  (1987).

\bibitem{RevModPhys.78.17}
P.~A. Lee, N. Nagaosa, and X.-G. Wen, Rev. Mod. Phys. {\bf 78},  17  (2006).

\bibitem{andi-dis}
P.~W. Anderson, J. Phys. Chem. Solids {\bf 11},  26  (1959).

\bibitem{doi:10.1021/ja800073m}
Y. Kamihara, T. Watanabe, M. Hirano, and H. Hosono, Journal of the American
  Chemical Society {\bf 130},  3296  (2008), pMID: 18293989.

\bibitem{PhysRevLett.101.026403}
L. Boeri, O.~V. Dolgov, and A.~A. Golubov, Phys. Rev. Lett. {\bf 101},  026403
  (2008).

\bibitem{0034-4885-74-12-124508}
P.~J. Hirschfeld, M.~M. Korshunov, and I.~I. Mazin, Reports on Progress in
  Physics {\bf 74},  124508  (2011).

\bibitem{doi:10.1146/annurev-conmatphys-020911-125055}
A.~V. Chubukov, Annual Review of Condensed Matter Physics {\bf 3},  57  (2012).

\bibitem{PhysRevLett.101.087004}
K. Kuroki {\it et~al.}, Phys. Rev. Lett. {\bf 101},  087004  (2008).

\bibitem{PhysRevB.80.180505}
R. Thomale, C. Platt, J. Hu, C. Honerkamp, and B.~A. Bernevig, Phys. Rev. B
  {\bf 80},  180505  (2009).

\bibitem{2009NJPh...11b5016G}
S. {Graser}, T. {Maier}, P. {Hirschfeld}, and D. {Scalapino}, New Journal of
  Physics {\bf 11},  025016  (2009).

\bibitem{PhysRevLett.101.076401}
Q. Si and E. Abrahams, Phys. Rev. Lett. {\bf 101},  076401  (2008).

\bibitem{PhysRevLett.101.206404}
K. Seo, B.~A. Bernevig, and J. Hu, Phys. Rev. Lett. {\bf 101},  206404  (2008).

\bibitem{RevModPhys.83.1589}
G.~R. Stewart, Rev. Mod. Phys. {\bf 83},  1589  (2011).

\bibitem{PhysRevLett.107.117001}
R. Thomale, C. Platt, W. Hanke, J. Hu, and B.~A. Bernevig, Phys. Rev. Lett.
  {\bf 107},  117001  (2011).

\bibitem{2012PhRvL.109h7001R}
J.-P. {Reid} {\it et~al.}, Physical Review Letters {\bf 109},  087001  (2012).

\bibitem{PhysRevB.89.064510}
A.~F. Wang {\it et~al.}, Phys. Rev. B {\bf 89},  064510  (2014).

\bibitem{PSSB:PSSB201600350}
C. Platt, G. Li, M. Fink, W. Hanke, and R. Thomale, physica status solidi (b)
  {\bf 254},  1600350  (2017), 1600350.

\bibitem{tommy}
T. B\"ohm {\it et~al.}, arXiv:1703.07749.

\bibitem{pome}
I. Pomeranchuk, JETP {\bf 20},  919  (1950).

\bibitem{rich}
R.~C. Richardson, Rev. Mod. Phys. {\bf 69},  683  (1997).

\bibitem{PhysRevB.57.8566}
K. Yang and S.~L. Sondhi, Phys. Rev. B {\bf 57},  8566  (1998).

\bibitem{Note1}
Orbital magnetic field effects of a nodal superconductor generically dominate
  over the Zeeman term at weak field strength for a field orientation
  perpendicular to the 2D plane. The Volovik effect~\cite {volo} then predicts
  a $\propto \protect \sqrt {H}$ scaling of the induced density of states, as
  opposed to the $\propto H$ scaling for the Zeeman term.

\bibitem{PhysRevLett.101.057003}
I.~I. Mazin, D.~J. Singh, M.~D. Johannes, and M.~H. Du, Phys. Rev. Lett. {\bf
  101},  057003  (2008).

\bibitem{PhysRevLett.102.047005}
F. Wang, H. Zhai, Y. Ran, A. Vishwanath, and D.-H. Lee, Phys. Rev. Lett. {\bf
  102},  047005  (2009).

\bibitem{PhysRevLett.106.187003}
R. Thomale, C. Platt, W. Hanke, and B.~A. Bernevig, Phys. Rev. Lett. {\bf 106},
   187003  (2011).

\bibitem{PhysRevB.79.094512}
V. Mishra {\it et~al.}, Phys. Rev. B {\bf 79},  094512  (2009).

\bibitem{1995JMP....36.1217L}
M.~H. {Lee}, Journal of Mathematical Physics {\bf 36},  1217  (1995).

\bibitem{2009arXiv0909.3653M}
M. Morales, arXiv:0909.3653.

\bibitem{abramowitz1972handbook}
M. Abramowitz and I.~A. Stegun, New York  361  (1972).

\bibitem{0953-8984-7-47-002}
T.~M. Rice and M. Sigrist, Journal of Physics: Condensed Matter {\bf 7},  L643
  (1995).

\bibitem{0295-5075-104-1-17013}
Q.~H. Wang {\it et~al.}, EPL (Europhysics Letters) {\bf 104},  17013  (2013).

\bibitem{elena}
E. Hassinger {\it et~al.}, Phys. Rev. X {\bf 7},  011032  (2017).

\bibitem{Hicks283}
C.~W. Hicks {\it et~al.}, Science {\bf 344},  283  (2014).

\bibitem{volo}
G.~E. Volovik, JETP Lett. {\bf 58},  457  (1993).

\end{thebibliography}






\end{document}